\documentclass[twocolumn,tighten,twocolappendix]{aastex701}

\usepackage{graphicx}
\usepackage{amsmath}
\usepackage{natbib}
\usepackage{amssymb}
\usepackage{ulem}
\usepackage{sidecap}
\usepackage{hyperref}
\usepackage{physics}

\shorttitle{Faint Radio and X-rays in TDE2025aarm}
\shortauthors{Matsumoto \& Piran}

\begin{document} 

\title{On the Faint Early-time Radio and X-ray Emissions in TDE2025aarm}

\author[0000-0002-9350-6793]{Tatsuya Matsumoto}
\affil{Department of Astronomy, School of Science, The University of Tokyo, Bunkyo-ku, Tokyo 113-0033, Japan}
\email[show]{matsumoto@astron.s.u-tokyo.ac.jp}
\author[0000-0002-7964-5420]{Tsvi Piran}
\affil{Racah Institute of Physics, The Hebrew University of Jerusalem, Jerusalem, 91904, Israel}
\email{}

\begin{abstract}
TDE2025aarm is a nearby tidal disruption event whose early radio and X-ray emissions are exceptionally faint compared with previously observed TDEs. We examine whether these weak signals can be explained within standard outflow and disk-emission scenarios. The radio detection at $15\,\rm GHz$ with $\sim10^{36}\,\rm erg\,s^{-1}$ around $40\,\rm days$ after discovery is inconsistent with synchrotron emission from a quasi-spherical disk wind for reasonable circum-nuclear densities and outflow velocities. Instead, the low luminosity and inferred self-absorbed spectrum imply a narrowly collimated outflow with a solid angle $\lesssim0.1\,\rm sr$, naturally identified with the unbound stellar debris. The X-ray emission is likewise unusually faint, with $L_{\rm X}\sim10^{39-40}\,\rm erg\,s^{-1}$ during the first few months. If interpreted as thermal emission from an obscured accretion disk, the inferred emitting area would correspond to an implausibly small X-ray-transparent region expected to vary on short dynamical timescales that are not observed. Alternatively, the same shock responsible for the radio emission can accelerate relativistic electrons that produce X-rays through synchrotron radiation and/or inverse-Compton scattering of optical/UV photons. Both mechanisms can explain the early faint X-ray emission, although their temporal evolution differs. Continued radio and X-ray monitoring of TDE2025aarm will provide a sensitive probe of the unbound debris, circum-nuclear medium, and high-energy emission mechanisms in optical TDEs.
\end{abstract}

\keywords{XXX}

\section{Introduction}
Tidal disruption events (TDEs) occur when a star is disrupted by a supermassive black hole (BH) in the center of a galaxy \citep{Hills1975,Rees1988}. They have been detected as transients in galactic nuclei across a wide range of wavelengths, from radio to gamma-rays \citep[e.g.,][]{Gezari2021}. An increasing number of TDEs are being discovered, and they serve as powerful probes of supermassive BHs, providing insights into accretion physics and BH demographics \citep[e.g.,][]{Yao+2023}. However, there are still several open questions regarding TDEs, for example, the emission mechanisms of optical/UV and X-rays, their connection to BH parameters, and the origin of the radio emission. One promising way to address these mysteries is to focus on nearby events and carry out detailed observations.

The recently discovered optical TDE2025aarm may offer such an opportunity. This event was discovered on 1st October 2025 by GOTO \citep{ONeill+2025_TNS} and classified as a nearby TDE at redshift $z\simeq0.0137$, corresponding to a luminosity distance as small as $d_{\rm L}\simeq60\,\rm Mpc$ \citep{ONeill+2025_TNS,Faris+2025_TNS,Newsome+2025_TNS}. In addition to extensive optical/UV observations, multiwavelength follow-up has been carried out. Several results have been reported on the Transient Name Server.\footnote{\texttt{https://www.wis-tns.org}} In the radio band, upper limits at $\lesssim8\,\rm GHz$ of $\simeq3\times10^{37}\,\rm erg\,s^{-1}$ were obtained on 31st October 2025 \citep{Sfaradi+2025_TNS}. At a higher frequency of 15GHz, \cite{Christy+2025_TNS} reported a detection with $\simeq2\times10^{36}\,\rm erg\,s^{-1}$ on 4th November 2025. X-rays were detected by Chandra with $\simeq3\times10^{39}\,\rm erg\,s^{-1}$ on 5th November 2025 \citep{Somalwar+2025_TNS}. \cite{Simongini+2026} reported Swift XRT observations with a luminosity of $\simeq10^{40}\,\rm erg\,s^{-1}$, derived from stacked data obtained at three epochs between October 2025 and February 2026. The most recent observation by \cite{Jaisawal+2026_TNS} revealed that the luminosity increased to $\simeq6\times10^{40}\,\rm erg\,s^{-1}$ in the observation conduced on 11th April 2026. No detection of very-high-energy gamma-rays has been reported \citep{Mohrmann+2025_TNS,Paneque+2025_TNS}.

\begin{figure}
    \centering
    \includegraphics[width=85mm,bb=0 0 277 190]{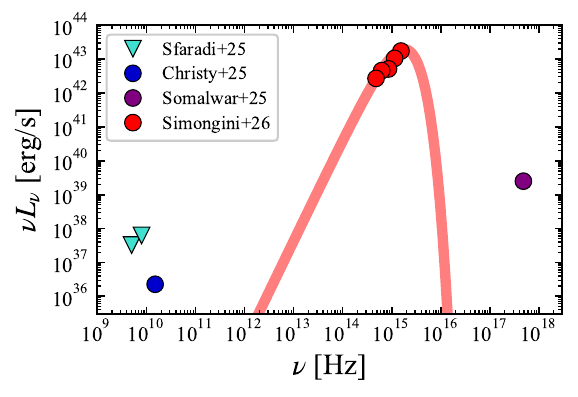}
    \caption{The spectral energy distribution of TDE2025aarm at $\simeq30-40\,\rm days$ after the optical discovery. Radio upper limits are from \cite{Sfaradi+2025_TNS}, the 15 GHz detection from \cite{Christy+2025_TNS}, the optical data from \cite{Simongini+2026}, and the X-ray measurements from \cite{Somalwar+2025_TNS}. Inverted triangles indicate upper limits, while circles denote detections. The pink curve represents the optical blackbody spectrum with $T_{\rm opt}=22000\,\rm K$ and $R_{\rm ph}=4\times10^{14}{\,\rm cm}$ \citep[see][]{Simongini+2026}.}
    \label{fig:25aarm_sed}
\end{figure}

Figure~\ref{fig:25aarm_sed} shows the spectral energy distribution constructed from the radio, optical, and X-ray observations summarized above, at $\simeq30-40\,\rm days$ after the optical discovery. The detected radio and X-ray luminosities are among the faintest observed in TDEs, and would not have been detectable at larger distances. In this Letter, we discuss the origin of these faint radio and X-ray emissions, which provides important insights into the underlying emission mechanisms.

\section{The Faint Radio Emission}
\label{sec:radio}
\cite{Christy+2025_TNS} reported the discovery of  faint radio emission at 15 GHz. The estimated luminosity $\simeq2\times10^{36}\,\rm erg\,s^{-1}$ places TDE2025aarm among the faintest radio detected TDEs to date. The spectral slope is positive within the radio band \citep{Christy+2025_TNS}. We interpret the radio signal as synchrotron emission from a forward shock driven by an outflow interacting with the surrounding medium, a scenario commonly invoked to explain radio flares in TDEs \citep[e.g.,][]{Krolik+2016,Alexander+2020}.\footnote{We assume that the outflow is Newtonian, and we do not consider any relativistic off-axis effect \citep[e.g.,][]{Matsumoto&Piran2023}.} In this standard picture, the outflow is slowed down by the circum-nuclear medium (CNM). The shock wave accelerates particles and enhances magnetic fields, resulting in synchrotron emission. 

A positive spectral slope is expected when the observed frequency $\nu_{15\rm GHz}=15\,\rm GHz$ falls in either $\nu_{15\rm GHz}<\nu_{\rm m}$ or $\nu_{\rm m}<\nu_{15\rm GHz}<\nu_{\rm a}$, where $\nu_{\rm m}$ and $\nu_{\rm a}$ are the characteristic synchrotron and synchrotron self absorption (SSA) frequencies, respectively \citep[e.g.,][]{Rybicki&Lightman1979,Sari+1998}. As the outflow is non-relativistic, the characteristic frequency is \citep{Matsumoto&Piran2021b}:\footnote{Throughout this paper, we assume the so-called deep Newtonian phase, in which the minimum Lorentz factor of shock-accelerated electrons is $\gamma_{\rm m}\simeq2$.}
\begin{align}
\nu_{\rm m}\simeq1.1\times10^{-2}{\,\rm GHz\,}\varepsilon_{\rm B,-2}^{1/2}n_{6}^{1/2}\left(\frac{\beta}{0.05}\right)\ ,
    \label{eq:nu_m}
\end{align}
where $\varepsilon_{\rm B}$, $n$, and $\beta$ are the fraction of post-shock energy in magnetic field, number density of the CNM, and outflow velocity normalized by the speed of light $c$, respectively. 
We adopt the notation $Q = Q_x10^x$ in cgs units, and use it hereafter unless otherwise specified.

\begin{figure}
    \centering
    \includegraphics[width=85mm,bb=0 0 277 213]{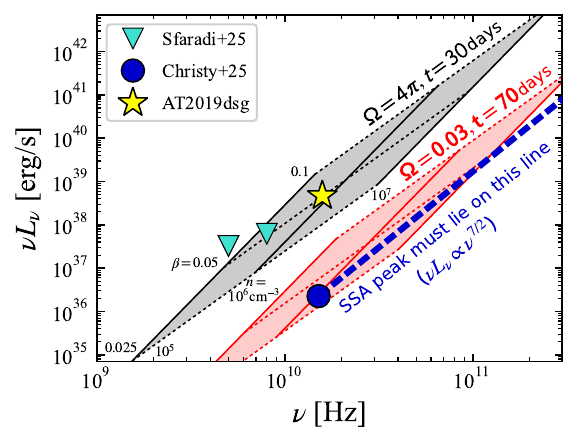}
    \caption{The possible locations of the SSA peak marked as shaded regions, obtained for the CNM density $n=10^{5-7}\,\rm cm^{-3}$ and outflow velocity $\beta=0.025-0.1$ (see  Eqs.~\ref{eq:nu_a} and \ref{eq:L_a}). Other parameters are fixed to $\varepsilon_{\rm e}=0.1$, $\varepsilon_{\rm B}=0.01$, and $p=2.5$. The solid lines depict the values for a constant density and varying velocity, while the dashed ones correspond to a constant-velocity with varying density.  The gray and pink regions correspond to opening angles of $\Omega=4\pi$ and $t=30\,\rm days$, and $0.03$ and $70\,\rm days$, respectively. The SSA peak of TDE2025aarm is expected to lie along the blue dashed line, which intersects the shaded regions only for a small opening angle. For comparison, the yellow star indicates the SSA peak of AT2019dsg at 42 days.}
    \label{fig:25aarm_ssa_peak}
\end{figure}

With typical values $\nu_{\rm m}$ falls below 15 GHz. We therefore expect the ordering
$\nu_{\rm m}\ll\nu_{15\rm GHz}<\nu_{\rm a}$. In this regime, SSA suppresses the
radio luminosity at $15\,\rm GHz$, and the spectrum follows
$\nu L_{\nu}\propto \nu^{7/2}$ up to $\nu_{\rm a}$. We estimate the SSA peak by extrapolating the
data along this $\nu L_\nu\propto \nu^{7/2}$ scaling, as indicated by the blue
dashed line in Fig.~\ref{fig:25aarm_ssa_peak}.

The potential SSA peak of TDE2025aarm is significantly offset from those of other TDEs. To illustrate this, we plot the SSA peak of AT2019dsg at $42\,\rm days$ after discovery \citep{Stein+2021} in Fig.~\ref{fig:25aarm_ssa_peak}. This event has an SSA peak at $\simeq16{\,\rm GHz}$, similar to that of TDE2025aarm, but with a much higher luminosity of $\simeq5\times10^{38}\,\rm erg\,s^{-1}$.

The SSA peak is closely linked to the physical properties of the radio-emitting outflow and is frequently used to estimate its size and energy under the assumption of equipartition between relativistic electrons and magnetic fields (the so-called equipartition method, \citealt{Chevalier1998,BarniolDuran+2013}).\footnote{The equipartition assumption is motivated by the fact that the total energy of the emitting region is minimized when the two are comparable, and increases rapidly away from equipartition \citep{BarniolDuran+2013}.} Here, we focus on the outflow velocity and the CNM density, and examine how these parameters determine the location of the SSA peak using Eqs.~(8) and (10) of \cite{Matsumoto&Piran2021b}:
\begin{align}
&\nu_{\rm a}\simeq21{\,\rm GHz\,}\bar{\varepsilon}_{\rm e,-1}^{\frac{2}{p+4}}\varepsilon_{\rm B,-2}^{\frac{p+2}{2(p+4)}}n_{6}^{\frac{p+6}{2(p+4)}}\left(\frac{\beta}{0.05}\right)^{\frac{p+8}{p+4}}\left(\frac{t}{30\rm day}\right)^{\frac{2}{p+4}}\ ,
    \label{eq:nu_a}\\
&\nu_{\rm a}L_{\nu_{\rm a}}\simeq1.1\times10^{39}{\,\rm erg\,s^{-1}\,}\bar{\varepsilon}_{\rm e,-1}^{\frac{7}{p+4}}\varepsilon_{\rm B,-2}^{\frac{3p+5}{2(p+4)}}n_{6}^{\frac{3p+19}{2(p+4)}}
	\nonumber\\
&\,\,\,\,\,\,\,\,\,\,\,\,\,\,\,\,\,\,\,\,\,\times\left(\frac{\beta}{0.05}\right)^{\frac{5p+34}{p+4}}\left(\frac{t}{30\rm day}\right)^{\frac{2p+15}{p+4}}\left(\frac{\Omega}{4\pi}\right)\ ,
    \label{eq:L_a}
\end{align}
where $\bar{\varepsilon}_{\rm e}=4\varepsilon_{\rm e}(p-2)/(p-1)$. $\varepsilon_{\rm e}$, $p$, and $\Omega$ are the fraction of post-shock energy in non-thermal electrons, the power-law index of the non-thermal electrons, and the solid angle of the outflow (in steradians), respectively. For the latter one, we adopt $p=2.5$ as a fiducial value unless otherwise noted. This choice is primarily motivated by observationally inferred values of $p=2-3$ \citep[e.g.,][]{Alexander+2016,Goodwin+2023b,Cendes+2024}. We confirmed that our conclusions remain essentially unchanged for values of $p$ other than 2.5. Our fiducial values of $\varepsilon_{\rm e}=0.1$ and $\varepsilon_{\rm e}=0.01$ are motivated by values commonly adopted in GRB afterglow modeling \citep[e.g.,][]{Panaitescu&Kumar2001b,Santana+2014}, as these parameters have not yet been systematically constrained for TDEs. We also assume a constant expansion velocity, so that the radius is given by $R=c\beta t$.

The shaded regions in Fig.~\ref{fig:25aarm_ssa_peak} indicate the regions of SSA peak locations for reasonable parameter ranges of $0.025\leq\beta\leq0.1$ and $10^5{\,\rm cm^{-3}}\leq n\leq10^7\,\rm cm^{-3}$. These values are suggested by radio-flare modeling of other TDEs \citep[e.g.,][]{Krolik+2016,Matsumoto&Piran2021b}. The gray region corresponds to a spherical outflow with $\Omega=4\pi$ and $t=30\,\rm days$ (i.e., launched around optical discovery). Remarkably, it does not overlap with the required locus of the SSA peak for TDE2025aarm, even when other parameters, except for $t$, are varied. For example, eliminating $\bar{\varepsilon}_{\rm e}$ from Eqs.~\eqref{eq:nu_a} and \eqref{eq:L_a}, we find that the SSA peak luminosity depends on frequency with $\nu_{\rm a}L_{\nu_{\rm a}}\propto \nu_{\rm a}^{7/2}$, which is the same scaling as the optically thick spectrum, $\nu L_\nu\propto \nu^{7/2}$. This implies that varying $\bar{\varepsilon}_{\rm e}$ mainly shifts the gray region along the SSA locus (blue dashed line), rather than toward it. A similar behavior is found when eliminating other parameters, $\nu_{\rm a}L_{\nu_{\rm a}}\propto \nu_{\rm a}^{3-4}$.

The SSA peak can intersect the required locus only if the outflow has a very small opening angle, $\Omega\ll 1$. The required opening angle, $\Omega\lesssim0.1$, is much smaller than that expected for a wind. However, such a narrowly collimated outflow arises naturally from the unbound debris \citep{Strubbe&Quataert2009,Krolik+2016,Yalinewich+2019b}. The unbound debris corresponds to roughly half of the stellar material, which is initially ejected with a velocity of $\beta\simeq0.05$, and is concentrated near the original orbital plane with $\Omega\sim10^{-3}-10^{-1}$. The pink shaded region corresponds to an example of the unbound debris case with $\Omega=0.03$ and $t=70\,\rm days$. As the unbound debris is ejected at the moment of disruption, its interaction with the CNM precedes the return of the most bound debris to the BH vicinity by $t_{\rm fb}\simeq40\,\rm days$. Therefore, the timescale appearing in Eqs.~\eqref{eq:nu_a} and \eqref{eq:L_a} should be replaced by $t+t_{\rm fb}\simeq30+40\,\rm days$, where $t$ is measured from the optical discovery. In addition, we consider a relatively large canonical value of $\beta=0.05$. 
The average value of the debris velocity is smaller. However, we use this value since at this early stage of the interaction between the outflow and the CNM only the tip of the unbound debris, which is moving much faster than the average, collides with the CNM \citep[e.g.,][]{Ryu+2020d,Matsumoto&Piran2021b}. Overall, the radio properties of TDE2025aarm favor a narrowly collimated outflow over a quasi-spherical one, making the unbound debris the most natural origin of the observed radio emission.

\section{The Faint X-ray Emission}
\label{sec:X}
\cite{Somalwar+2025_TNS} reported Chandra X-rays detection at $\simeq40\,\rm days$ after the discovery. Due to the low flux, the spectral shape is not well constrained, but it is consistent with a single power-law with a photon index of $\simeq2$, corresponding to an unabsorbed luminosity of $L_{\rm X}\simeq3\times10^{39}\,\rm erg\,s^{-1}$. \cite{Simongini+2026} reported Swift XRT X-ray detections at $\simeq60$ and $120\,\rm days$, with an indication of increasing flux. The spectrum can be fitted by either a power-law with a photon index of $\simeq2$, which is consistent with \cite{Somalwar+2025_TNS}, or as a blackbody 
with a temperature of $k_{\rm B}T\simeq0.4{\,\rm keV}$ ($k_{\rm B}$ is the Boltzmann constant). The inferred unabsorbed luminosity is roughly $L_{\rm X}\simeq(6-12)\times10^{39}\,\rm erg\,s^{-1}$. More recently \cite{Jaisawal+2026_TNS} found that the luminosity increased to $\simeq6\times10^{40}\,\rm erg\,s^{-1}$ at $\simeq190\,\rm days$.

About $\simeq40\,\%$ of optically discovered TDEs exhibit X-ray emission. Their spectra are consistent with blackbody one with temperature of $k_{\rm B}T\sim0.1{\,\rm keV}$, and their luminosities in the early phase spans a wide range of $L_{\rm X}\sim10^{41-44}{\,\rm erg\,s^{-1}}$ \citep{Guolo+2024b}. In this context, TDE2025aarm stands out as exhibiting one of the lowest X-ray luminosities ever observed among TDEs. Again this detection is possible due to its relatively small distance, which allows such faint emission to be detected. 

A widely discussed scenario for TDEs' X-ray emission is the obscured accretion disk model. In this picture, an accretion disk is already formed by the time of optical discovery, but it is enshrouded either by an outflow (\citealt{Strubbe&Quataert2009,Metzger&Stone2016,Dai+2018,Lu&Bonnerot2020,Thomsen+2022}, but see \citealt{Matsumoto&Piran2021}) or by an envelope surrounding the BH \citep{Loeb&Ulmer1997,Krolik+2016, Metzger2022b}. The surrounding material does not completely cover the disk; Depending on the viewing angle, X-rays can be significantly obscured along some lines of sight, while escaping with little attenuation along others \citep{Krolik+2016,Dai+2018}. The covering fraction of the solid angle can be roughly constrained by estimating the size of the X-ray emitting region as $\sim(R_{\rm X}/R_{\rm g})^2$, where $R_{\rm g}=GM_\bullet/c^2$ is the gravitational radius ($G$ and $M_\bullet$ are the gravitational constant and BH mass, respectively), which sets the characteristic size of the (total) emitting region. Indeed, for faint X-ray events, the size of the inferred emission region can be as small as $R_{\rm X}\lesssim R_{\rm g}$, suggesting a nearly complete covering of the disk \citep{Gezari2021,Yao+2022b,Guolo+2024b}.

For TDE2025aarm, the low X-ray luminosity implies that the emission region must be extremely compact. This casts doubt on the obscured disk scenario. The size of the X-ray emitting region can be estimated as 
\begin{align}
R_{\rm X}\sim\left(\frac{L_{\rm X}}{4\pi \sigma_{\rm SB}T^4}\right)^{1/2}\simeq4\times10^9{\,\rm cm\,}L_{\rm X,40}^{1/2}\left(\frac{k_{\rm B}T}{0.1\,\rm keV}\right)^{-2}\ ,
    \label{eq:Rx}
\end{align}
where $\sigma_{\rm SB}$ is the Stefan-Boltzmann constant. This is much smaller than the gravitational radius of the BH, $R_{\rm g}\simeq2\times10^{11}\,M_{\bullet,6}$. However, such a small ``hole'', namely a channel with reduced optical depth in the obscuring material is unlikely to be stable and should instead be highly variable, leading to rapid time variability in the observed X-ray flux. The accretion disk and surrounding environment are turbulent, making it difficult to maintain such a small X-ray-transparent opening over an extended period. A natural variable timescale is the dynamical timescale, given by $\sim R_{\rm X}/c_{\rm s}\simeq4\times10^3{\,\rm s}$, where we adopt a sound speed of $c_{\rm s}\simeq(k_{\rm B}T_{\rm opt}/m_{\rm p})^{1/2}\simeq10{\,\rm km\,s^{-1}\,}T_{\rm opt,4}^{1/2}$ ($m_{\rm p}$ is the proton mass), and the temperature is the one of the optical photosphere. Therefore, it appears very unlikely that most of the disk X-ray emission is obscured while allowing only a tiny fraction to escape through a small opening without a significant temporal variability.

We note that the size estimate based on the Stefan-Boltzmann law (Eq.~\ref{eq:Rx}) is a simplified approach and is subject to systematic uncertainties. One such uncertainty is the color correction in the disk atmosphere, which relates the observed color temperature to the effective temperature through $T_{\rm c}=f_{\rm c}T_{\rm eff}$ \citep[e.g.,][]{Shimura&Takahara1995b}. Although $f_{\rm c}$ is not well established for TDE disks, values commonly adopted for active galactic nuclei and X-ray binaries typically modify the inferred effective temperature by at most a factor of $f_{\rm c}\simeq2$ \citep[e.g.,][]{Davis&ElAbd2019}, corresponding to a factor of $\simeq4$ in the inferred emitting size. Such a correction does not qualitatively alter our conclusion. In addition, even in the absence of an obscuring envelope, the observed X-ray luminosity from a bare disk is reduced for an edge-on observer. However, reproducing the persistently faint X-ray emission over the first few months would still require a finely tuned inclination angle, similar to the argument presented above.

\begin{figure}
    \centering
    \includegraphics[width=85mm,bb=0 0 277 195]{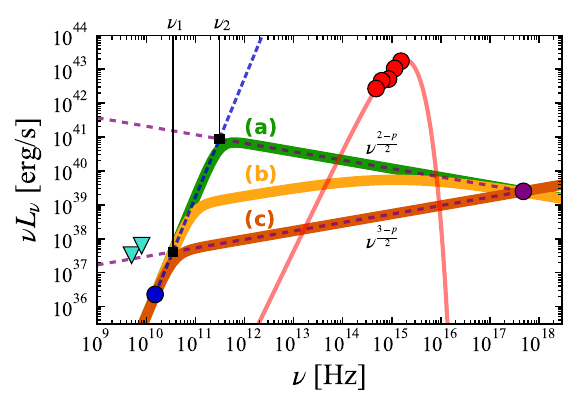}
    \includegraphics[width=85mm,bb=0 0 285 163]{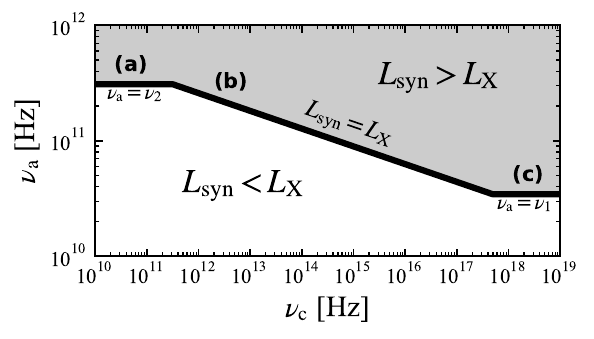}
    \caption{({\bf Top}) The same as Fig.~\ref{fig:25aarm_sed} but possible synchrotron spectra are presented, which can produce the observed X-rays. Cases (a)-(c) correspond to different ordering of the key frequencies $\nu_{\rm a}$, $\nu_{\rm c}$, $\nu_{\rm X}$, and $\nu_2$. ({\bf Bottom}) The configuration of the key frequencies $\nu_{\rm a}$ and $\nu_{\rm c}$. Synchrotron emission reproduces the observed X-ray luminosity only when both frequencies are located on the solid line. Above the line (gray region), the predicted luminosity exceeds the observed value, while below it the emission is insufficient.}
    \label{fig:25aarm_sed_obs2}
\end{figure}

We propose an alternative scenario.  The faint X-rays arise from non-thermal relativistic electrons accelerated by the same shock that accelerates the lower energy non-thermal electrons responsible to the radio emission.
The non-thermal electrons produce X-rays via synchrotron\footnote{A reverse shock may also accelerate non-thermal electrons and produce synchrotron X-rays. However, unlike the forward shock case, its parameters are poorly constrained by the current data, and we therefore do not discuss this possibility further.} and via inverse Compton (IC) of optical photons. 

We consider, first,  synchrotron emission. The broadband spectrum must contain one or two spectral breaks between the radio and X-ray bands. The position of these breaks depends on the self-absorption frequency and on the synchrotron cooling frequency, above which electrons lose their energy over the dynamical timescale:
\begin{align}
\nu_{\rm c}&\simeq50{\,\rm GHz\,}\varepsilon_{\rm B,-2}^{-3/2}n_{6}^{-3/2}\left(\frac{\beta}{0.05}\right)^{-3}\left(\frac{t}{70\,\rm day}\right)^{-2}\ ,
	\label{eq:nu_c}
\end{align}
where we use the expressions given in \cite{Matsumoto2025}, and the values for $\beta$ and $t$ are motivated by the unbound debris scenario. It is also useful to define two characteristic frequencies corresponding to the intersections between the SSA locus and the low-frequency extrapolation of the X-ray power-law spectrum with slopes $\nu^{(3-p)/2}$ and $\nu^{(2-p)/2}$, 
which are expected for synchrotron emission by slow and fast cooling electrons \citep[e.g.,][]{Sari+1998},
respectively:
\begin{align}
\nu_{1}&=\left(\frac{L_{\rm X}}{L_{\rm 15GHz}}\nu_{\rm 15GHz}^{7/2}\nu_{\rm X}^{(p-3)/2}\right)^{\frac{2}{p+4}}\simeq35{\,\rm GHz}\ ,\\
\nu_{2}&=\left(\frac{L_{\rm X}}{L_{\rm 15GHz}}\nu_{\rm 15GHz}^{7/2}\nu_{\rm X}^{(p-2)/2}\right)^{\frac{2}{p+5}}\simeq310{\,\rm GHz}\ ,
\end{align}
where $L_{\rm 15GHz}$ is the observed luminosity at $15\,\rm GHz$. The numerical values are evaluated for the fiducial choice of $p=2.5$.

As shown in the top panel of Fig.~\ref{fig:25aarm_sed_obs2}, depending on the relative ordering of the four key frequencies, $\nu_{\rm a}$, $\nu_{\rm c}$, $\nu_{\rm X}$, and $\nu_2$, three synchrotron configurations can account for the observed X-ray luminosity:
\begin{description}
\item[(a) $\nu_{\rm c}<\nu_{\rm a}=\nu_{2}<\nu_{\rm X}$] 
The X-rays are produced by fast-cooling electrons, giving a photon index of $(p+2)/2$ at the whole range between $\nu_{\rm a} $ and $\nu_{\rm X}$. In this case, the self-absorption frequency must satisfy $\nu_{\rm a}=\nu_2$. If $\nu_{\rm a}$ is lower (higher) than $\nu_2$, the predicted synchrotron X-ray luminosity becomes smaller (larger) than the observed value.

\item[(b) $\nu_{\rm a}<\nu_{2}<\nu_{\rm c}<\nu_{\rm X}$] 
As in case (a), the X-rays are produced by fast-cooling electrons. But in this configuration,  there is a spectral break at $\nu_{\rm c} $ which is above the self-absorption frequency. The latter must satisfy
\begin{align}
\nu_{\rm a}=\nu_{2}\left(\frac{\nu_{\rm c}} {\nu_{2}}\right)^{-1/(p+4)}\ >\nu_1\ ,
\end{align}
for the synchrotron X-ray luminosity to reproduce the observed value.

\item[(c) $\nu_{\rm a}<\nu_{2}<\nu_{\rm X}<\nu_{\rm c}$] 
The X-ray-emitting electrons are in the slow-cooling regime, giving a photon index of $(p+1)/2$ between $\nu_{\rm X}$ and $\nu_{\rm a}$ that must satisfy $\nu_{\rm a}=\nu_1$. However, this possibility may not be realized unless the magnetic field is extremely weak, $\varepsilon_{\rm B}\lesssim10^{-6}$ (see Eq.~\ref{eq:nu_c}).
\end{description}
At present, all three scenarios remain viable, although we consider case (c) less likely because it requires an extreme value of $\varepsilon_{\rm B}$. Identifying one of the spectral breaks would provide a constraint on the other key frequency as shown in the bottom panel of Fig.~\ref{fig:25aarm_sed_obs2}.

\begin{figure}
    \centering
    \includegraphics[width=85mm,bb=0 0 285 215]{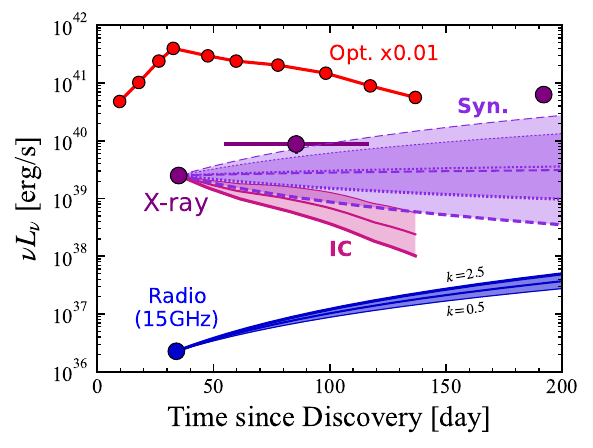}
    \caption{Multiwavelength light curves of TDE2025aarm. The purple shaded regions represent the possible X-ray light curves predicted by our models. The dashed, dotted, solid purple curves correspond to the synchrotron (slow- and fast-cooling) and IC scenarios, respectively. The width of each curve reflects the assumed slope of the CNM density profile, $k=0.5$ (thin), 1.5 (medium), and 2.5 (thick). We also show the expected radio light curves at 15 GHz (blue), which lie in the self-absorbed regime, $\nu L_\nu\propto n^{-1/4}R^2$. The optical light curve (red) is taken from \cite{Simongini+2026}.}
    \label{fig:25aarm_lc}
\end{figure}

The temporal evolution of the synchrotron X-ray luminosity is given, for $\Omega=const$,  by $\propto t^{\frac{8-k(p+2)}{4}}$ for the fast cooling cases (a) and (b), and $L_{\rm X}\propto t^{\frac{12-k(p+5)}{4}}$ in the other case. Depending on the CNM slope $k$, the X-ray luminosity can either rise or decline with time as shown in Fig.~\ref{fig:25aarm_lc}. In particular, for a shallow CNM profile $k\simeq0.5$, the synchrotron emission may explain also the long-term X-ray light curve.\footnote{The X-ray brightening observed by \cite{Jaisawal+2026_TNS} at $\simeq190\,\rm days$ may instead be explained by the emergence of the accretion-disk emission as the reprocessing envelope becomes optically thin, or by delayed disk formation \citep[e.g.,][]{Gezari+2017b}.} An exact evaluation of the light curve has to take into account the detailed hydrodynamics of the outflow. The luminosity will increase if $\Omega $ increases, and it will decrease as the outflow slows down. A joint analysis with the radio emission will be crucial for constraining this scenario.

In addition to synchrotron the accelerated electron may upscatter optical/UV photons via IC as discussed in the context of supernovae \citep{Fransson1982,Bjornsson&Fransson2004,Chevalier&Fransson2006}. Since TDEs show bright and long lasting optical/UV emission and are often accompanied by radio emission, IC up-scattering is expected to occur naturally 
\citep[e.g.,][]{Kumar+2013}. 
Optical/UV photons originating from the inner photosphere, with a characteristic size of $\sim10^{15}\,\rm cm$, propagate to the radio emitting region and are partially up-scattered via the IC process to X-rays, $\nu_{\rm X}\sim \gamma^2 \nu_{\rm opt}$, where $\gamma$ is the electron Lorentz factor and $\nu_{\rm opt}$ is the frequency of the seed optical/UV photons \citep[e.g.,][]{Rybicki&Lightman1979}. We remark that the physical origin of the optical photons is not important in this scenario. Our IC mechanism only requires the observed optical/UV photon field as seed photons, regardless of how it is produced.

We are interested in electrons with Lorentz factors $\gamma=(\nu_{\rm X}/\nu_{\rm opt})^{1/2}\simeq30\,\nu_{\rm X,18}^{1/2}\nu_{\rm opt,15}^{-1/2}$. The fraction of photons scattered by electrons with $\gamma$ is given by the optical depth \citep[e.g.,][]{Rybicki&Lightman1979}:
\begin{align}
\tau_{\rm IC}(\gamma)&\simeq\sigma_{\rm T}n_{\rm e}(\gamma)\Delta R
    \label{eq:tau}\\
&\simeq4\times10^{-6}\,n_{\rm e,tot,6}\Delta R_{15}\left(\frac{\gamma}{30}\right)^{1-p}\ .
    \nonumber
\end{align}
Here $\sigma_{\rm T}$ is the Thomson cross section. Assuming a power-law electron distribution with an index of $p$, the number density is given by $n_{\rm e}(\gamma)\sim n_{\rm e,tot}\gamma^{1-p}$, where $n_{\rm e,tot}$ is the total number density of non-thermal electrons. The radial width of the shocked radio-emitting region is typically a tenth of the outflow radius, $\Delta R\sim0.1\,R$. Since each scattered photon gains energy by a factor of $\gamma^2$, the IC luminosity can be estimated as
\begin{align}
L_{\rm IC}&\sim \gamma^2\tau_{\rm IC}(\gamma) L_{\rm opt}\left(\frac{\Omega}{4\pi}\right)
    \label{eq:L_IC}\\
&\simeq 9\times10^{38}{\,\rm erg\,s^{-1}\,}n_{\rm e,tot,6}\Delta R_{15}\Omega_{-1}
    \nonumber\\
&\,\,\,\,\,\,\,\,\,\times\left(\frac{\gamma}{30}\right)^{3-p}\left(\frac{L_{\rm opt}}{3\times10^{43}{\,\rm erg\,s^{-1}}}\right)\ ,
    \nonumber
\end{align}
which is consistent with the initial observed faint X-ray luminosity. Here we include the geometrical factor $\Omega/4\pi$, motivated by the small opening angle of the radio emitting outflow.

The IC X-ray luminosity depends on both the luminosity of the seed optical/UV photons and the optical depth ($\propto n_{\rm e,tot}R$). Since the former is relatively well constrained by optical observation, the X-ray light curve can be used to probe the CNM density profile if the IC scattering is the dominant emission mechanism. When the CNM density follows a power-law profile with index $k$ and the outflow expands at a constant velocity, the column density evolves as $n_{\rm e,tot}R\propto t^{1-k}$. This introduces an additional time dependence in $L_{\rm X}$ beyond that of $L_{\rm opt}(t)$. Figure~\ref{fig:25aarm_lc} illustrates the expected temporal evolution of the IC X-ray light curve. Given the gradual decline of the optical/UV light curve, the X-ray luminosity is expected to fade slowly. Therefore, IC emission alone may not be sufficient to explain the entire X-ray light curve unless the outflow undergoes a significant lateral expansion and $\Omega$ increases. The observed X-ray spectrum is consistent with the IC emission, in which the spectrum follows a power-law with a photon index of $(p+1)/2\simeq1.8$ for $p=2.5$.

\section{Summary}
\label{sec:summary}
TDE2025aarm is one of the nearest TDEs discovered to date. The reported early-time observations revealed surprisingly faint radio and X-ray emissions, which we have investigated in this Letter. We propose that both are produced by non-thermal relativistic electrons that are accelerated in shocks between an outflow and the CNM. For the radio emission, we find that the expected SSA peak luminosity is too low to be explained by a spherical outflow such as a wind. Instead, it is consistent with a narrowly collimated outflow with a solid angle of $\lesssim0.1$ and $\beta \simeq 0.05$. The unbound debris is the most natural origin of such an outflow. 

The X-ray emission is also among the faintest observed in TDEs. A naive estimate suggests an extremely compact emitting region of $\sim10^9\,\rm cm$. In the widely discussed obscured disk scenario, such a small region would correspond to a hole through which disk X-rays can escape without significant attenuation. However, given the turbulent environment around the BH, such a small region is expected to exhibit rapid variability on timescales of $\sim10^{3-4}\,\rm s$, which is not observed.

We therefore propose an alternative scenario for the faint X-ray emission. The X-rays can be generated by non-thermal electrons accelerated by the same shock that accelerated the radio-emitting electrons. The specific emission mechanism could be  either synchrotron emission from the same electron population responsible for the radio emission, or IC scattering of optical/UV photons from the central photosphere by the synchrotron-emitting electrons. Both mechanisms are inevitable, but the former one could be more promising to explain the long-timescale evolution since the latter predicts a declining light curve on a shorter timescale. However, our prediction partly relies on some simplifications such as the constant velocity and opening angle of the outflow. Realistically the outflow decelerates slowly and spreads laterally, which effects on the light curve evolution.  Detailed observations of these faint, previously undetected emission components will provide important insights into the nature of TDEs.



\begin{acknowledgements}
The authors thank the Yukawa Institute for Theoretical Physics at Kyoto University. Discussions during the YITP long-term workshop YITP-T-26-02 on ``Multi-Messenger Astrophysics in the Dynamic Universe'' were useful to complete this work. T.M. is also grateful to Kenta Hotokezaka for stimulating discussions. This research is supported by JSPS KAKENHI (grant No. 24K17088), an advanced ERC grant Multijets, and the Simons SCEECS collaboration.
\end{acknowledgements}

\bibliographystyle{aasjournalv7}
\bibliography{reference_matsumoto}

\end{document}